\documentstyle[aps,epsfig]{revtex}

\begin{document}

\draft

\title{The dynamical structure factor in disordered systems}

\author{V. Mart\'{\i}n-Mayor, G. Parisi and P. Verrocchio}

\address{Dipartimento di Fisica,
Universit\`a di Roma ``La Sapienza'',
Piazzale Aldo Moro 2, 00185 Roma (Italy) \\
INFN sezione di Roma - INFM unit\`a di Roma}

\date{\today}
\maketitle

\begin{abstract}
We study the spectral width as a function of the external momentum for
the dynamical structure factor of a disordered harmonic solid,
considered as a toy model for supercooled liquids and glasses. Both in
the context of single-link coherent potential approximation and of a
single-defect approximation, two different regimes are clearly
identified: if the density of states at zero energy is zero, the
Rayleigh $p^4$ law is recovered for small momentum. On the contrary,
if the disorder induces a non vanishing density of states at zero
energy, a linear behaviour is obtained. The dynamical structure factor
is numerically calculated in lattices as large as $96^3$, and
satisfactorily agrees with the analytical computations.
\end{abstract}


The spectrum of vibrational excitations of supercooled liquids and
glasses is attracting a great deal of attention from the experimental
side~\cite{ESPERIMENTI}, the numerical simulations~\cite{SIMULAZIONI}
and also from the analytical point of
view~\cite{TEORIA,GOTZE}. Vibrational excitations on the GHz range
develop in the supercooled liquid, by the same temperatures where
anomalous behaviour of the specific-heat is found. Some rather
substance-independent features in this vibrational spectrum are
found. For example, the vibrational density of states presents an
excess respect to the usual Debye behaviour
($g(\omega)\propto\omega^2$) known as the {\em Boson Peak}. The
dynamical structure-factor, $S(p,\omega)$, reveals well-defined
sound-like peaks for wave-lengths not much greater than the
inter-particle distance. Moreover, the speed of this high-frequency
sound is close to the one of the low-frequency typical one. For fixed
external momentum, $p$, the width of the spectral peak grows as $p^2$,
which has recently been recovered in the mode coupling approximation
(see~\cite{GOTZE}). However, in Ref.~\cite{SCHIRMACHER}, a simpler
model of a disordered three-dimensional harmonic solid was studied.
There it was claimed that if some spring-constants are allowed to have
a small negative value, (but not so much that negative-energy modes
appear) a boson peak develops.  In the context of the CPA
approximation~\cite{KIRKPATRICK}, the usual $p^4$ width of
$S(p,\omega)$ arisen from Rayleigh-scattering is reported.  It was,
however, noticed that a rather different scaling appeared at the
characteristic frequencies of the Boson-Peak.  Furthermore, in
Ref.~\cite{MONTAGNA}, it was shown a $p^2$ broadening for a
one-dimensional disordered harmonic solid model.  In this paper, we
want to investigate the problem, both by numerical and analytical
means. Our main finding will be that the Rayleigh-scattering $p^4$
broadening holds, unless the solid becomes {\em unstable}. That is, if
the dynamical-matrix, to be defined later, has an {\em extensive}
number of negative eigenvalues, the system do present sound-like peaks
in its $S(p,\omega)$, but with a width proportional to $p$, at least
for very small values of $p$. Otherwise, the standard $p^4$ behaviour
is to be expected.

To be more specific, our dynamical matrix in $D$ dimensions is given by
\begin{eqnarray}
H&=&\frac{1}{2} \sum_{xy} \phi_x {\cal H}_{xy} \phi_y\label{HAMILTONIANA}\,,\\ 
{\cal H}_{xy}&=&\sum_{\mu=1}^D
\frac{1+\alpha_{y,\mu}}{2}\left(\delta_{xy}-
\delta_{x,y+\hat\mu}\right) \nonumber \\
&+&\frac{1+\alpha_{y-\hat\mu,\mu}}{2}\left(\delta_{xy}-\delta_{x,y-\hat\mu}\right)\label{DYNMATRIX}\, ,
\end{eqnarray}
where $\hat\mu$ is the lattice unit-vector in the $\mu$ direction.
Notice that we cannot separate the longitudinal and transverse modes
since, for the sake of simplicity, our matrix has no internal 
indices as in Ref.~\cite{SCHIRMACHER}.
The $\alpha_{y,\mu}$ is the random part of the spring constant that joins
the sites $y$ and $y+\hat{\mu}$.
In a finite lattice,
periodic boundary conditions are applied. The dynamical
matrix, verifies the constraint related with traslational
symmetry: the vector of all equal components is 
eigenvector with zero eigenvalue. 
For a harmonic solid, the object that we will study is the
dynamical structure function, $S(p,E)$, in the energy rather
than in the frequency-domain:
\begin{equation}
\label{RESOLVE}
S(p,E)=
-\frac{1}{\pi} {\mathrm {Im}} \lim_{\epsilon\to 0} 
\overline{\langle p|\frac{1}{E+ i\epsilon - {\cal H}}|p\rangle}\, . 
\end{equation}
In the above expression, the bra-ket notation has been used and $|p\rangle$
stands for a normalized plane-wave of momentum $p$. 
As usual, the overline represents the average over the random variables 
$\alpha$. We choose to work in the energy, rather than in the
frequency domain, since we mainly want to consider the case where a
significant fraction of the spectrum is negative, and so the
transformation $E=\omega^2$ is no longer well defined.

We study the case where the $\alpha_{y,\mu}$ are uncorrelated,
random variables whose probability distribution is
\begin{equation}
p(\alpha)=(1-\rho)\delta(\alpha)+\rho f(\alpha), 0\le\rho\le 1\,. 
\label{PALPHA}
\end{equation} 
In the above equation, $\rho$ is the probability of finding one {\em
defect}, while $f$ is a continuous probability function, that we take
flat between $\lambda$ and $0$ ($\lambda<0$). Notice that the
disorder of Ref.~\cite{SCHIRMACHER} is recovered, by taking $\rho=1$ and $f$
Gaussian. The rationale for choosing this kind of disorder is
that a spring of an unusually large
negative spring-constant induces a negative energy eigenvalue of the
dynamical matrix $\cal H$. In fact, if one keeps in
Eq.(\ref{DYNMATRIX}) only the spring connecting sites $1$ and $2$,
corresponding to the large negative defect, the eigenvectors are
easily shown to be $\phi_x=(\delta_{x,1}+\delta_{x,2})/\sqrt{2}$ and
$(\delta_{x,1}-\delta_{x,2})/\sqrt{2}$. However, assuming that the
surrounding springs have their ordered value, one finds that the
Hamiltonian (\ref{HAMILTONIANA}) is negative for the displacement
configuration $(\delta_{x,1}-\delta_{x,2})/\sqrt{2}$ if the
spring-constant verifies $\alpha<-(D+\frac{1}{2})$. Therefore, one has
a contribution of order $\rho$ to the density of states over the
negative spectrum. Moreover, if we assume that some of the surrounding
spring have very small positive values, a contribution of higher-order
in $\rho$ to the density of states is generated, no matter how small
is the negative value of the spring constant, $\alpha+1$. Therefore,
if the probability of $\alpha<-1.0$ is non-vanishing, the
hybridization of the localized field-configuration described above,
with the plane-waves eigenstates will non-trivially modify the
eigenvectors of ${\cal H}$.
To study this, we shall perform a single-defect
calculation of the resolvent.

In this way, we learn that the order $\rho$
threshold for the presence of negative eigenvalues is not $\alpha=-D-1/2$,
as roughly shown in the introduction, but $-D$. This threshold separate
two well defined scaling limits for the width of the $S(q,E)$ at small $q$:
\begin{itemize}
\item
If the density of states is null at zero energy the imaginary 
part of the self-energy $\Sigma$ is proportional to $p^{D+2}$ 
(Rayleigh scattering, that yields a spectral width $\propto p^{D+1}$
in the frequency domain).
\item
When an extensive number of negative eigenvalues is present, 
$\Sigma \propto p^2$ (or $p$ for the width in the frequency domain).
Of course, the $p^{D+2}$ contribution will be still present but sub-dominant
at low momentum. A crossover might be visible, depending on the strength
of the disorder. 
\end{itemize}

The dynamical matrix in the presence of an unique defect of amplitude
$\alpha$ can be split in two terms:
\begin{equation}
{\cal H}_{xy} = {\cal H}_{xy}^0 + {\cal R}_{xy}
\label{perturba}
\end{equation}
where the pure crystal matrix ${\cal H}^0$ has plane waves eigenvectors
with eigenvalues
\begin{equation}
E_0(p) = \sum_{\nu} (1-\cos\,p_{\nu})\,.
\end{equation}
The perturbation, hence, connects the two sites $y^0$ and $y^0+\nu$:
\begin{eqnarray}
{\cal R}_{xy} & \equiv &\alpha^{y^0;\nu} |y^0;\mu\rangle\langle y^0;\mu|\,, \\
|y^0;\mu\rangle & \equiv & \frac{\delta_{x,y^0} - \delta_{x,y^0+\nu}}{\sqrt{2}}\,.
\end{eqnarray}
The propagator can be written as
\begin{equation}
\frac{1}{z-{\cal H}} = \frac{1}{z-{\cal H}^0} +\frac{1}{z-{\cal H}^0} 
{\cal T}  \frac{1}{z-{\cal H}^0}\,, 
\end{equation}
where the resummation of the harmonic series gives
\begin{equation}
{\cal T} \equiv |y^0;\nu\rangle \frac{\alpha}{1- \alpha a(z)} \langle y^0;\nu | \,,
\label{UNTERMINO}
\end{equation}
with
\begin{equation}
a(z) \equiv \frac{1}{D} \int \frac{d^D q}{(2 \pi)^D}
\frac{E_0(q)}{z-E_0(q)}\,. 
\label{ADIZ}
\end{equation}

Notice that the
correction term in Eq.(\ref{UNTERMINO}), has an isolated singularity for the value of $z$
satisfying $1-\alpha a(z)=0$. This value decrease monotonically from
$z=0$, for $\alpha=-D$, and will be called $z_\alpha$. This isolated
singularity, correspond to the ground state of the dynamical 
matrix, the residue being $\Psi_F(x)\Psi^*_F(y)$. Therefore, one
obtain the wave-function for this eigenvalue
\begin{equation}
\Psi_F(x)  \propto \int \frac{d^D q}{(2 \pi)^D}
\frac{1-e^{i q_{\nu}}}{z_{\alpha}-E_0(q)} e^{i q x}\, ,  
\end{equation}
that has a localization length of order $|z_\alpha|^{-1/2}$. We thus
see, that unless $\alpha$ was exceedingly close to the critical value
$-D$, the eigenvector is strongly localized around the defect. The
single-defect approximation to the self-energy,
Eq.(\ref{SDselfenergy}), amounts to consider that each defect only
contributes to its own localized eigenvector, and that no other defect
is within its localization length.
Let us turn back to the original problem, with an extensive
number of defects, and repeat the above calculation, neglecting
all terms which contain two different defects. The matrix ${\cal T}$
is now a sum of terms like the one in Eq.(\ref{UNTERMINO}). If
we now perform the average over the $\alpha$'s and apply Dyson 
resummation, the self-energy part of the propagator turns out to be
\begin{equation}
\Sigma(z,p) = \rho E_0(p) \int d \alpha f(\alpha) 
\frac{\alpha}{1-\alpha a(z)}\,,
\label{SDselfenergy}
\end{equation}
at first order in $\rho$ (the defect interactions will generate
the order $\rho^2$ and higher order corrections to the single defect
result).

The width of the $S(q,E)$ is simply given by the value of the
imaginary part of the self-energy at the peak, whose position can be
obtained from the real part of the self-energy,
\hbox{$E^{\mathrm{max}}(p)\approx E_0(p)+ \mathrm{Re} \Sigma(E_0(p),p)$}. Notice that
our self energy is proportional to $E_0(p)$ so that we are basically
getting a finite renormalization of the speed of sound. In order
to estimate the imaginary part of the self energy, it is useful
to realize that for small positive values of the energy, one has
\begin{equation}
a(E+i\epsilon) \sim -\frac{1}{D} - i \frac{E^{D/2}}
{2^{D/2} \pi^{D/2-1} \Gamma(D/2)}\,.
\end{equation}
Therefore, if the probability density $f(\alpha)$ do not allow
$\alpha$ to be smaller than $-D$, the only imaginary term in
Eq.(\ref{SDselfenergy}) comes from $a(E+i\epsilon)$, and it is of
order $p^2E^{3/2}$, yielding a value of order $p^5$ at the peak. On
the other hand, if $\alpha$ can be smaller than $-D$, an imaginary
part of order $p^2 f(-D)$ arises from the pole.

Led by the functional form of the self-energy in Eq.(\ref{SDselfenergy}), one
can consider self-energies of the form $f(E)E_0(p)$, also when $\rho$ is not
small, and the single defect approximation no longer holds. This is
the idea lying under the well-known CPA approximation~\cite{KIRKPATRICK},
where one sets $\Sigma(z,p)=(\Gamma(z)-1)E_0(p)$. A self consistency
equation can be readily written:
\begin{equation}
\overline{\frac{1+\alpha-\Gamma(z)}{\Gamma(z)-a \left(z/\Gamma(z) \right)
\left(1+\alpha-\Gamma(z) \right) }} = 0\,.
\label{CPAEQ}
\end{equation}
It is clear that the
width of the $S(q,E)$ critically depends on the value of
$\Gamma(0+i\epsilon)$.  For the flat distribution of $\alpha$
introduced in Eq.(\ref{PALPHA}), one can solve Eq.(\ref{CPAEQ}) in the limit
of small $\rho$, as $\Gamma(0)=1 + b \rho + {\cal O}(\rho^2)$
obtaining
\begin{equation}
b = D \left( 1 -D \log{\frac{D}{|\lambda+D|}} +
i \pi \frac{D}{\lambda} \theta(D+\lambda) \right)\,,
\label{SINGLE}
\end{equation}
where $\theta(x)=1$ for $x<0$ and zero otherwise.  It is easy to check
that the single-defect result is exactly 
the same as in Eq.(\ref{SINGLE}).

Fixing from now on $D=3$,
Eq.(\ref{CPAEQ}) in the general case, can be numerically solved. 
We choose to write it as a {\em
fixed-point} equation, and solve it recursively. The only tricky part,
is the evaluation of $a(z)$ defined in Eq.(\ref{ADIZ}), that we make
by calculating the unperturbed density of states by a Monte Carlo
simulation. In this way, we are able to obtain estimates of
$\Gamma(z)$ with an accuracy of $10^{-4}$. We find that in the CPA
approximation, there are the same two regimes as in the single-defect
computation, separated by a critical line, whose coordinates are shown
in table \ref{TAVOLACALDA}. The appearance of the $p^5$ regime
coincides, and it is due to, the vanishing of the negative-energy
spectrum. However, as explained in the introduction, one expects
rather that what vanish is the order $\rho$ contribution to the
density of states in this region. In fact, we expect a non-vanishing
density of states all the way down to $\lambda=-1$, where no negative
spring exists. It is remarkable that when $\rho=1$, 
the CPA approximation keeps a non-vanishing fraction of negative eigenvalues
up to $\lambda=-1.2$, quite close to the correct value. 

\begin{figure}[t!]
\begin{center}
\epsfig{file=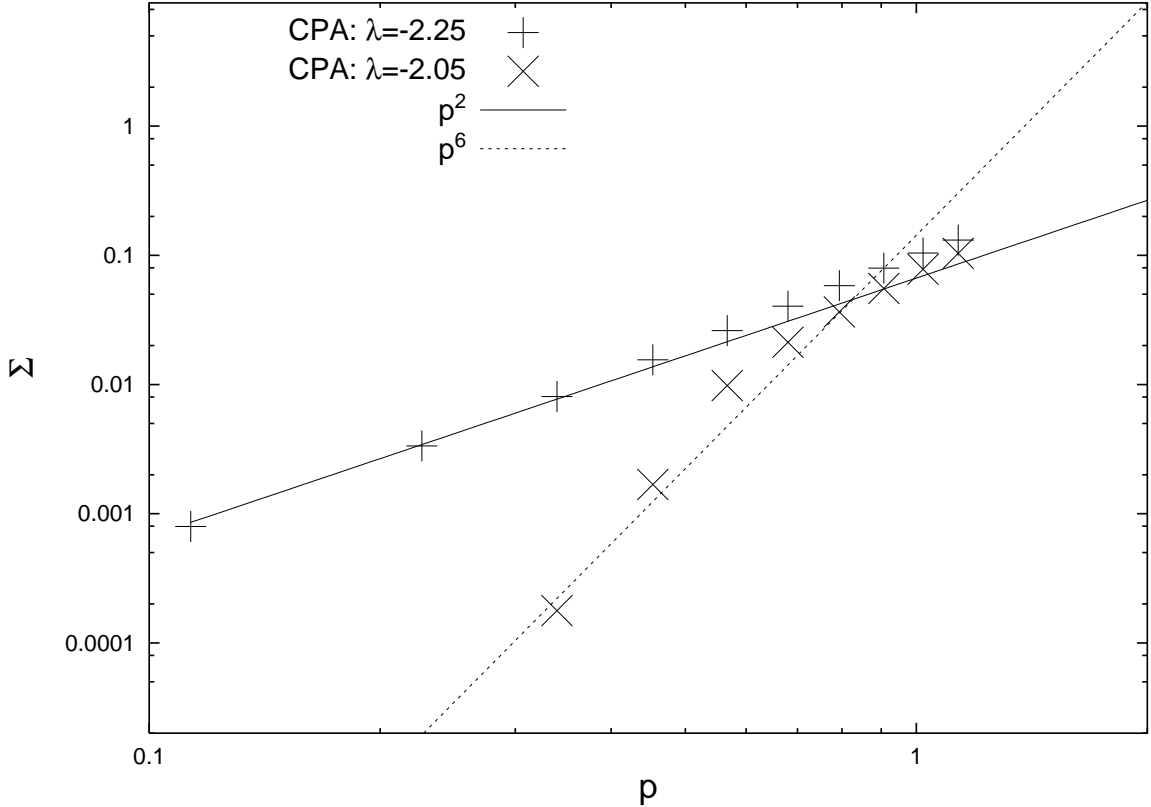,width=11cm,angle=270}
\end{center}
\caption{The spectral width $\Sigma$ (see Eq.(\ref{FIT})) as a function
of the external momentum in the CPA approximation, for a density of
defects $\rho =0.1$}
\label{CPAFIG}
\end{figure}

\begin{figure}[t!]
\begin{center}
\epsfig{file=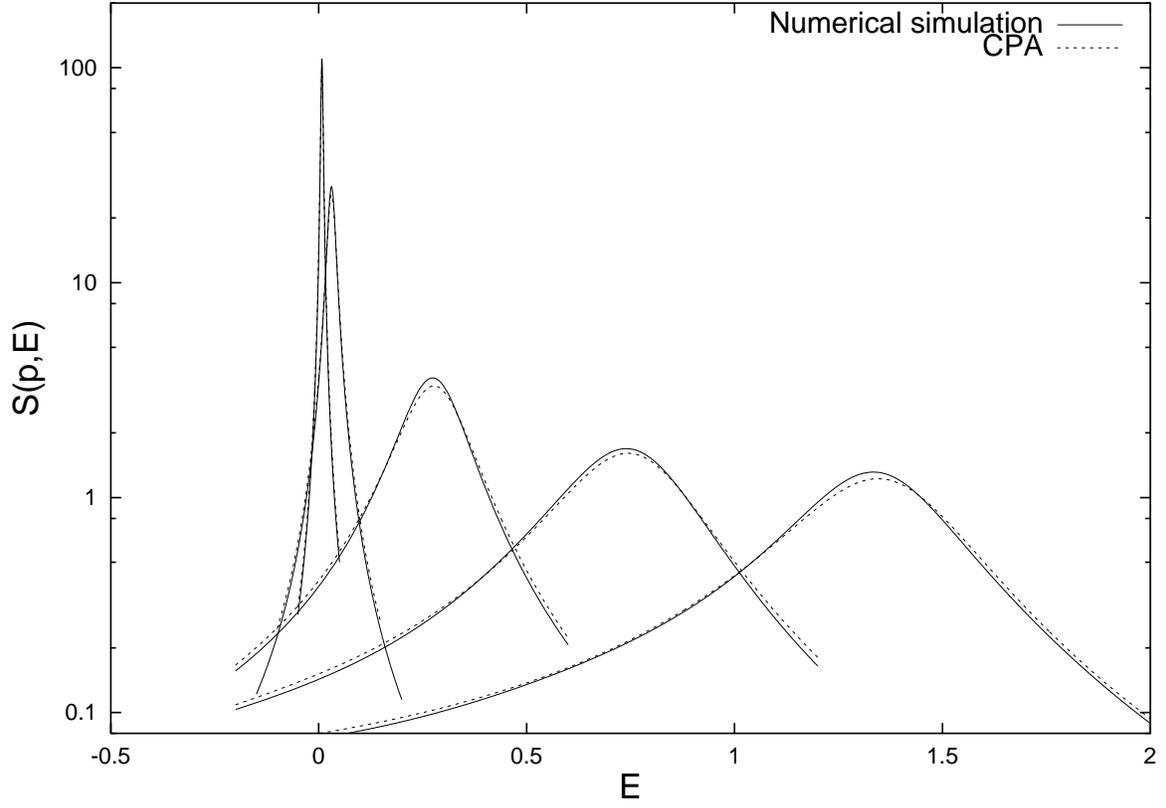,width=11cm,angle=270}
\end{center}
\caption{The $S(q,E)$ function, for $p=2\sqrt{3}\pi\,n/96$ for
$n=8,10,12,14$, from left to right, both in the CPA approximation and
in the numerical calculation, for $\rho =0.1$ and $\lambda=-10.0$.}
\label{CPANUML10}
\end{figure}

\begin{figure}[t!]
\begin{center}
\epsfig{file=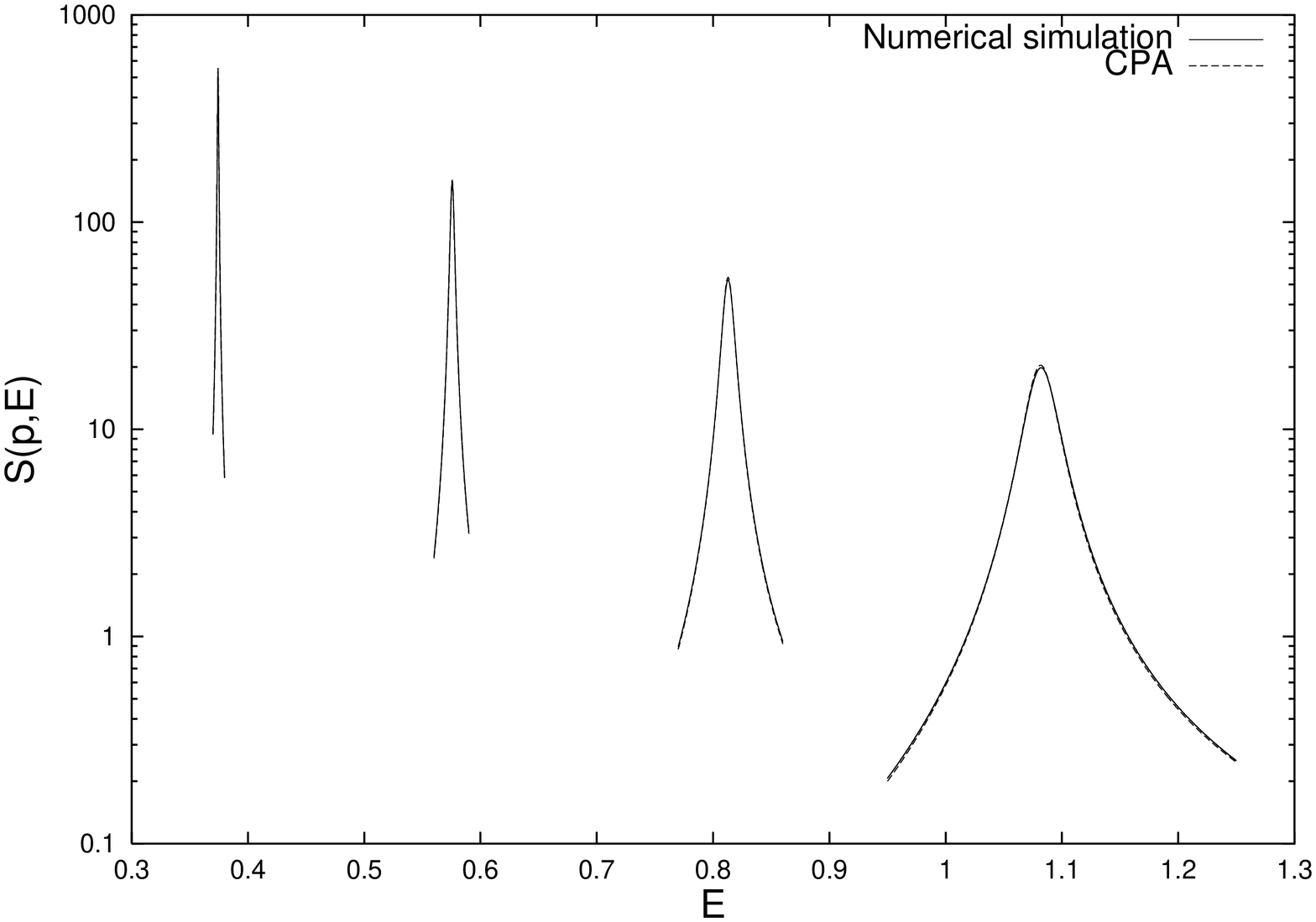,width=11cm,angle=270}
\end{center}
\caption{The $S(q,E)$ function, for $p=2\sqrt{3}\pi\,n/96$ for
$n=1,2,6,10,14$, from left to right, both
in the CPA approximation and in the numerical calculation, for
$\rho =0.1$ and $\lambda=-1.0$.}
\label{CPANUML1}
\end{figure}

The agreement between the CPA approximation and the results from
numerical simulations for the $S(E,p)$ turns out to be better 
than $5\%$ in the two extreme 
cases (see figs. \ref{CPANUML10},\ref{CPANUML1}).
That is the one where there is no negative springs ($\lambda=-1$) and
the one where the spring constant can be very negative
($\lambda=-10$).  Close to the CPA critical line (in the unstable
crystal side) the agreement is still quite good. On the (CPA) stable
border side, a scaling between $p^6$ and $p^5$ is found, depending on the
density of defects, $\rho$. For example,
with the value of the density of defects $\rho=0.1$, in fig.~\ref{CPAFIG}) the
transition between the two regimes found by the CPA approximation
at $\lambda=-2.15$ is very evident.

\begin{figure}[t!]
\begin{center}
\epsfig{file=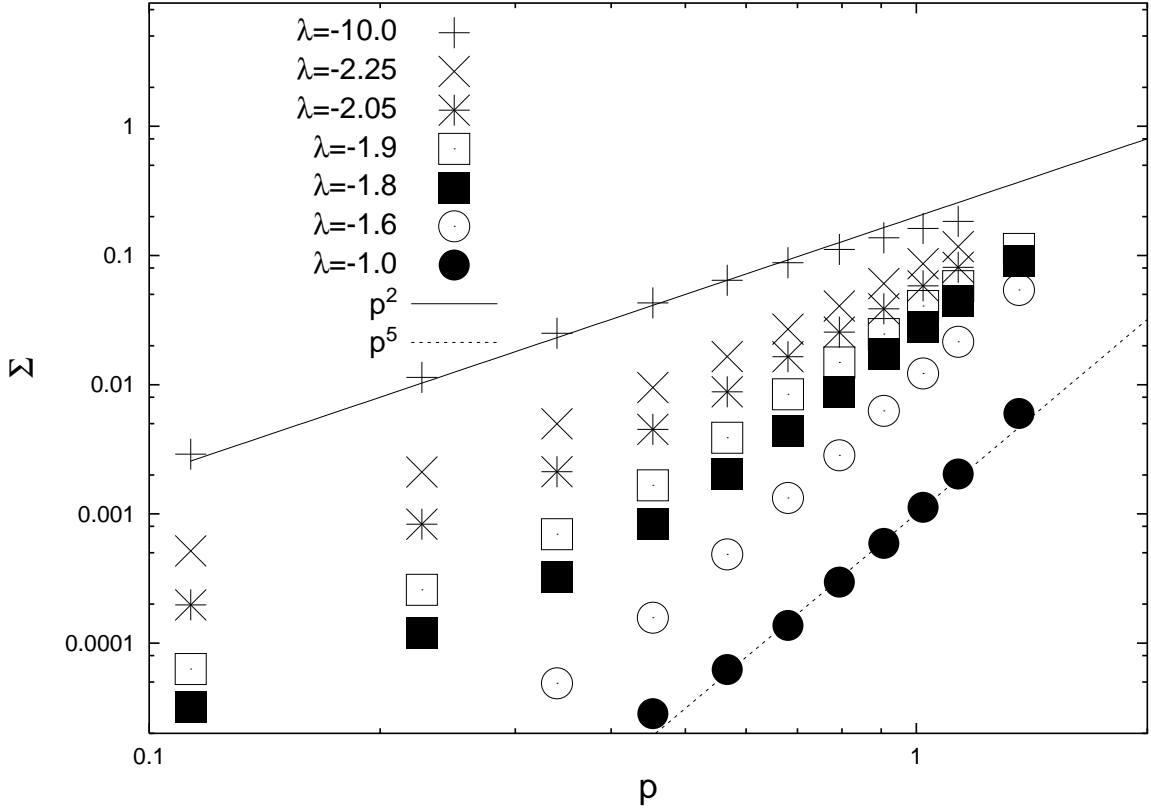,width=11cm,angle=270}
\end{center}
\caption{$\Sigma$ as a function of the external momentum in the
numerical simulations, for a density of defects $\rho =0.1$. The errors are
smaller than the data points.}
\label{SIMTUTTIFIG}
\end{figure}

We numerically computed the structure factor $S(E,q)$ for our model
utilizing the method of moments~\cite{MOMENTI}, which allows to study
the statistical properties of the eigenvalues and the eigenvectors of
large dynamical matrices, avoiding their diagonalization.  This method
is a clever modification of the Lanczos method and it shares the same
weakness, namely the lack of orthogonality when a too large number of
moments is computed.  Another limitation is the necessity of setting a
finite value of $\epsilon$ in Eq.(\ref{RESOLVE}). A reasonable value
for $\epsilon$ is the mean distance, between eigenstates, which is
roughly given by $(2D-\lambda)/V$, where $V$ is the lattice volume
(that is $10^{-5}$ in our $96^3$ lattice). If the width of the peak is
comparable with $\epsilon$, the results will be definitely affected by
finite-size effects. The limitation related with the number of
moments, is not serious for the central part of the $S(q,E)$ curve,
but can be rather strong if one wants to calculate the tails of the
distribution. In practice, we have used $30$ moments for $10$
different disorder realizations, finding very satisfying results,
unless the peak height approaches values of order $10^5$,
when finite size effects turns out to be important.
We fit our results to the Breit-Wigner 
form:
\begin{equation}
S(E,q) = \frac{{\cal N}}{(E - E_0)^2+\Sigma^2}\,,
\label{FIT}
\end{equation}
which satisfactorily describes the peak in all cases, although
usually overestimates the tails of the distribution.
The position of the peak $E_0$ is linear in $p^2$ as expected.
More importantly,
the results from numerical simulations (fig. \ref{SIMTUTTIFIG})
show very clearly that the
critical line previously discussed is an artifact of the
CPA approximation and a real system actually becomes unstable when
there is an extensive number of negative springs, no matter how small
is the fraction of them, as expected from the simple analytic
considerations sketched previously.
For larger values of $p$ the $p^5$ contribution is
no longer negligible and a complicated intermediate scaling appears.

In this work, we used the fact that both the single-defect
approximation and the CPA approximation~\cite{KIRKPATRICK} allow to
write the self-energy of a disordered harmonic solid, for small $E$
and $p$, as $f(E)E_0(p)$. An unavoidable consequence of this
functional form is that violations of the Rayleigh $p^5$ scaling of the
width of the peaks of the $S(p,E)$ are intrinsically tied with the
appearance of a negative energy spectrum ({\em i. e.} solid
instability). We have argued that an extensive number of negative
eigenvalues should appear, as soon as the spring-constants are allowed
to be negative. Our numerical calculations on $96^3$ disordered
lattices confirm the above picture. For wave-lengths in the range
$[a, 10a]$ ($a$ being the lattice constant), a
crossover regime may come out, depending on the strength of the disorder,
from the competition between the $p^2$ term and
$p^5$ term in the self-energy.
Therefore, the vibrational excitations
of the disordered solid can look qualitatively similar to the
finite temperature Instantaneous Normal Modes of supercooled 
liquids and glasses, where negative eigenvalues are always present.

We gratefully acknowledge interesting discussions with A. Gonzalez,
M. M\'ezard, G. Ruocco and G. Viliani.  V.M.M. is a M.E.C. fellow and
has been partially suported by CYCyT(AEN97-1708 and AEN99-1693). Our
numerical computations have been carried out on the Kalix2 pentium
cluster of the University of Cagliari.

\begin{center}
\begin{table}
\caption{The critical line $\lambda_{\mathrm c}(\rho)$ in the CPA approximation.}
\label{TAVOLACALDA}
\begin{tabular}{|c|c|c|c|c|c|c|c|}
$\rho$ & 0.0001 &0.1   &0.2  &0.4   &0.6  &
0.8   & 1.0    \\ \hline 
$-\lambda$ & 2.995(5) & 2.15(5) & 1.84(1) & 1.55(2) &
 1.38(1) &  1.28(1) & 1.22(1) \\ 
\end{tabular}
\end{table}
\end{center}

\end{document}